\documentclass[
    aps,
    prx,
    reprint,
    longbibliography,
    nofootinbib,
    superscriptaddress,
    floatfix
]{revtex4-2}

\usepackage{amsmath,amssymb,amsfonts,latexsym}
\usepackage{mathtools}
\usepackage{bm}            
\usepackage{mathrsfs}      
\usepackage{tensor}

\usepackage{graphicx}
\usepackage{dcolumn}       
\usepackage{appendix}
\usepackage{relsize}
\usepackage{parallel}
\usepackage{enumitem}

\usepackage[usenames,dvipsnames]{xcolor}
\usepackage{hyperref}
\hypersetup{
    colorlinks=true,
    linkcolor=blue,
    filecolor=magenta,
    urlcolor=cyan,
    pdfpagemode=UseNone,
}
\begin{document}

\title{Coupling between gravitational and electromagnetic perturbations on Kerr Spacetime}

\author{Fawzi Aly
}
\thanks{Corresponding author}
\email{mabbasal[AT]buffalo.edu}
\author{Dejan Stojkovic
}
\email{ds77[AT]buffalo.edu}
\affiliation{HEPCOS, Physics Department, SUNY at Buffalo, Buffalo, New York, USA}

\begin{abstract}
We extend our previous Schwarzschild metric-based  studies of gravitational--electromagnetic (GEM) coupling to rotating black holes by working directly in a curvature-based Newman--Penrose/Teukolsky framework on Kerr spacetime. Within a minimally coupled Einstein--Maxwell system, we derive explicit quadratic electromagnetic source terms for the spin-$-2$ Teukolsky equation, providing a foundation for future numerical studies of GEM interactions in the framework of black-hole spectroscopy. Moreover, we give order-of-magnitude arguments showing that GEM quadratic quasinormal modes (QQNMs) can become relevant in a range of charged and magnetized astrophysical scenarios. Finally, we show through a brief dilaton-theory example that the GEM QQNM spectrum is sensitive to how gravity couples to electromagnetism, thereby providing a model-based way to test minimal coupling and to constrain hidden $U(1)$ sectors with gravitational-wave observations.
\end{abstract}

\maketitle

\section*{Introduction}
Studies of nonlinear effects have become a focus of the recent progress in gravitational-wave theory and phenomenology, providing deep insights into the intricate dynamics of strong-field regimes. Numerous studies have presented compelling evidence of nonlinear imprints in the ringdown phase of numerical waveforms simulating binary black hole (BBH) mergers \cite{NonlinearitiesCaltechMitman,Cheung_Nonlinearities_Berti_Grouup,NearHorizonNonlinearities,QQNMS_New_Filter,Greedy_fit_London_2014,Aly_2023}. Perturbative second-order calculations now play a critical role in both ringdown analyses \cite{Redondo_Yuste_2024,spin_denpdence_Hengrui_Zhu,Cheung_2024_paper,Baibhav_2023_agnostic,Ma_2024,Absorption_induced_mode_excitation,Lagos_2023_GreenFunctionAnalysis_Quadratic_Diracdelta,Ripley_1,Ripley_2} and modeling of extreme-mass-ratio inspiral (EMRI) waveforms \cite{Pound_2020,pound_2017,spiers2023_second_teukolsky_kerr,spiers2024_secondorder_perturbations_schwarzschildspacetime,Angelica_self_Force}. Nonlinearities are anticipated to have significant implications for the analysis of future high signal-to-noise ratio detections with Advanced LIGO, next-generation ground-based detectors such as the Einstein Telescope and Cosmic Explorer, and the upcoming LISA mission \cite{ET_Punturo_2010,ET_ScienceCase_2020,CE_Reitze_2019,CE_Evans_2021,LISA_AmaroSeoane_2017,Chamberlain_2017,From_LISA_Einstein,pectroscopy_of_massive,LISA_}, thereby shedding further light on strong-gravity physics.

In the context of BBH mergers, studies mostly focus on nonlinearities arising from gravity--gravity (GG) couplings, while largely ignoring matter--gravity couplings. That reflects the expectation that gravitational dynamics dominate high-energy astrophysical phenomena. Nevertheless, magnetized or charged systems, with their strong electromagnetic fields, exhibit a richer phenomenology and are natural targets for multi-messenger astronomy \cite{multi_messenger,Abbott_2017__,Margalit_2017,Branchesi_2016}. In such environments, additional nonlinearities can arise from matter--gravity interactions, namely gravity--electromagnetism (GEM) coupling. The question now is: what electromagnetic energy scale is required for GEM coupling to play a non-negligible role?

On the theoretical side, examining GEM effects offers a valuable framework for testing alternative gravity models and the principle of minimal coupling, and for exploring a broader class of applications. Nevertheless, GEM coupling could also contribute to the evolving landscape of multi-messenger astronomy, particularly following the detection of the GW170817 event \cite{abbott2017gw170817,Abbott_2017__}, which provided the first observation of gravitational waves (GWs) and electromagnetic radiation across the electromagnetic spectrum. With advances in observational instruments, many other multi-messenger sources could be observed in the future \cite{Gupta, rate_NSBH_merger}. Hence, from a phenomenological perspective, a natural step forward is to explore how these fields interact and influence each other. If GWs and electromagnetic radiation can leave mutual imprints on each other, different radiation channels will provide complementary and overlapping information about the same astrophysical event, offering new opportunities to probe and refine our understanding of the strong-field regime.

In this work, we focus on the GEM sector and build on previous analysis of GEM coupling on Schwarzschild backgrounds \cite{Fawzi_EM_idealdipole,aly2024nonlinearitieselectromagneticgravitationalmode}. We take the next step by extending this framework to the more astrophysically relevant case of rotating black holes. Our primary goal is to explore the structure of GEM interactions on Kerr spacetime, deriving the corresponding quadratic sources for the gravitational Teukolsky equation and casting them in a form suitable for practical applications. In addition, we present several plausibility arguments for the significance of GEM interactions across different astrophysical scenarios, namely neutron star--black hole (NSBH) EMRIs, binary neutron star (BNS) mergers as well as charged BBH mergers. Moreover, we discuss how GEM coupling is theory-dependent and illustrate, through a brief example, how this dependence can be used to test the minimal coupling principle.

The structure of this paper is as follows. In Sec.~\ref{GEM in Kerr spacetime} we formulate GEM coupling on Kerr spacetime within the Newman--Penrose formalism and derive explicit expressions for the quadratic electromagnetic source of the gravitational Teukolsky equation. In Sec.~\ref{sec:GEM_coupling} we briefly recap the associated GEM QQNMs and their relation to linear gravitational and electromagnetic QNMs. In Sec.~\ref{Discussion} we assess the plausibility of observing GEM effects in several astrophysically motivated scenarios. We summarize our findings and outline future directions in Sec.~\ref{Conclusion}.

\section{GEM in Kerr spacetime}\label{GEM in Kerr spacetime}
According to the procedure outlined in \cite{aly2024nonlinearitieselectromagneticgravitationalmode}, analyzing the GEM coupling for the gravitational sector involves first solving for the electromagnetic perturbation  \(\phi^{(1)}_{i}\) (with \(i = 0, 1, 2\)) and then using the resulting electromagnetic stress-energy tensor as input for the gravitational perturbations equations. Specific calculations may vary depending on the scenario. For EMRIs, the first-order electromagnetic perturbation equations are sourced by a non-zero four-current in addition to the initial conditions. On the other hand, for ringdown analysis, the remnant is treated as a perturbed stationary background (e.g., a Kerr black hole in this work), where only initial data is relevant.

In either case, the particular solution is obtained through convoluting the Green’s function (for the Teukolsky kernel \cite{spiers2023_second_teukolsky_kerr}) of the electromagnetic perturbation with the effective source. The resulting solution is then used to construct the effective source for the gravitational perturbation via the electromagnetic stress-energy tensor \cite{aly2024nonlinearitieselectromagneticgravitationalmode, Fawzi_EM_idealdipole}, allowing for the analysis of GEM modes at the first level of mixing\footnote{The first level of mixing refers to the hierarchy between the energy scales of “pure” gravitational phenomena and electromagnetic ones. See \cite{aly2024nonlinearitieselectromagneticgravitationalmode,Fawzi_EM_idealdipole} for further discussion.}. This is a concise sketch of the procedure. Here we only focus on the perturbed, minimally coupled Einstein--Maxwell system on the Kerr background.

The study of both gravitational and electromagnetic perturbations in the Petrov type D spacetime is carried out in a null tetrad as outlined in \cite{Teukolsky1_1973}. Teukolsky’s work unified these perturbations into a single master partial differential equation:
\begin{equation}
   {}_{s} \mathcal{T} \chi^{(k)}_{s} = S_{\mathrm{eff}},
\end{equation}
where \( \chi^{(k)}_{s} \) denotes a generic perturbation of either the first order \( k = 1 \), second order \( k = 2 \), or the first level of mixing \( k \equiv \text{GEM} \), while \( S_{\mathrm{eff}} \) represents the corresponding effective source term. This master equation unifies the treatment of gravitational, electromagnetic, and scalar field perturbations within a common framework.  In what follows, we will denote GEM modes as \(\psi^{\textit{(GEM)}}_{j}\) for the gravitational sector and \(\phi^{\textit{(GEM)}}_{i}\) for the electromagnetic sector, while second-order gravitational perturbations will be denoted by \(\psi^{(2)}_{j}\).

On the gravitational side, our primary focus is on the radiative component \( \psi^{\text{(GEM)}}_{4} \), governed by the equation:  
\begin{equation}
\begin{gathered}
    {}_{-2} \mathcal{T} \psi^{\text{(GEM)}}_{4} = 8 \pi \Sigma \rho^4 T^{\textit{EM}}_{4},
\end{gathered}
\end{equation}
where \( \rho = r - i a \cos \theta \) and \( \Sigma = \rho \rho^{*} \), working in the Kinnersley tetrad $(l^\mu,n^\mu,m^\mu,m^{* \mu})$ and Boyer--Lindquist coordinates $(t,r,\theta,\varphi)$. The source could be decomposed into
\begin{equation}
\begin{gathered}
    8 \pi \Sigma \rho^4 T^{\textit{EM}}_{4} = \mathcal{A} \, T^{\mathrm{EM}}_{nm^*} + \mathcal{B} \, T^{\mathrm{EM}}_{m^*m^*} + \mathcal{C} \, T^{\mathrm{EM}}_{nn},
\end{gathered}
\end{equation}
where \( \mathcal{A}, \, \mathcal{B}, \, \mathcal{C} \) correspond to differential operators acting on the electromagnetic stress-energy components and are defined in \cite{Teukolsky1_1973} [Eq. 2.15], up to a factor of \( 8 \pi \Sigma \rho^4 \).  \( T^{\mathrm{EM}}_{nm^*}, \, T^{\mathrm{EM}}_{m^*m^*}, \, T^{\mathrm{EM}}_{nn} \) represent contractions of the electromagnetic stress-energy tensor with null tetrad vectors. 

As shown in \cite{aly2024nonlinearitieselectromagneticgravitationalmode}, \( T_{4}^{\mathrm{EM}} \) can be rearranged to decompose into two pieces: \( T^{\text{lin}}_{4}\left( \phi^{(1)}_{i}, J^{\text{Matter}} \right) \) and \( T^{\text{quad}}_{4}\left( \phi^{(1)}_{i} \phi^{(1)}_{j} \right) \). The first term is primarily relevant in EMRI problems and controls GEM contributions that are linear in the electromagnetic perturbations, while the second term is quadratic. As discussed in this work and further analyzed in \cite{aly2024nonlinearitieselectromagneticgravitationalmode,Fawzi_EM_idealdipole}, minimal coupling partially controls how GEM effects manifest as linear and quadratic in electromagnetic perturbations.

To obtain \( T^{\textit{EM}}_{4} \), we need to construct the electromagnetic stress-energy tensor \( T^{\mathrm{EM}}_{\mu \nu} \) from the first-order perturbation of the Maxwell scalars. This requires solving for \( \phi_{2}^{(1)} \) and \( \phi^{(1)}_{0} \), and subsequently constructing \( \phi^{(1)}_{1} \) in a manner similar to that described in \cite{chandrasekhar_1983}:

\begin{equation}\label{phi_0,2 inhomo}
\begin{gathered}
   {}_{1} \mathcal{T} \phi_{0}^{(1)}= 4\pi \Sigma  J^{\textit{Matter}}_{0}, \\
       {}_{-1} \mathcal{T} \phi_{2}^{(1)}=4 \pi \Sigma \rho^2 J^{\textit{Matter}}_{2} .
        \end{gathered}
\end{equation}
\begin{equation}
\begin{gathered}
\phi_0 = F_{\mu \nu} l^\mu m^\nu, \\
\phi_1 = \frac{1}{2} F_{\mu \nu}\left(l^\mu n^\nu + m^{* \mu} m^\nu\right),\\
\Phi_{2}^{(1)} = F_{\mu \nu} m^{* \mu} n^\nu,\\
\end{gathered}
\end{equation}
After a simple rescaling of $\phi^{(1)}_{i} \equiv \Phi^{(1)}_{i} 2^{-i/2} \rho^{i}$, we can write the relevant components of $T^{\mathrm{EM}}$ as 
\begin{equation}
\begin{gathered}
     T^{\mathrm{EM}}_{nm^*}= \frac{\sqrt{2}}{8 \pi} \frac{\Phi_{2}^{(1)} \Phi_{1}^{(1)*}}{ \Sigma \rho} \,, \quad\quad 
     T^{\mathrm{EM}}_{m^*m^*}=\frac{2}{8 \pi} \frac{\Phi_{2}^{(1)} \Phi_{0}^{(1)*}}{ \rho^2}, \\
    T^{\mathrm{EM}}_{nn}=\frac{1}{8 \pi }  \frac{|\Phi_{2}^{(1)}|^{2}}{\Sigma^2},\\
\end{gathered}
\end{equation}
where \( |\Phi_{2}^{(1)}|^{2} = \Phi_{2}^{(1)} \Phi_{2}^{(1)\, *} \), $\Delta \equiv r^2 - 2 M r + a^2, \qquad R^2 \equiv r^2 + a^2$. We can express the contribution of electromagnetic source terms to gravitational perturbations as follows:
\begin{widetext}
\begin{equation}
\begin{gathered}
\mathcal{A} \, T^{\mathrm{EM}}_{nm^*}= - \frac{2 i \rho \sin \theta (a^2 + R^2 \csc^2 \theta)}{\rho^*} \frac{\partial^2 \Phi_{2}^{(1)} \Phi_{1}^{(1)*}}{\partial t \partial \varphi}  + \frac{2 i \csc \theta}{\rho^*} 
    \frac{\partial \Phi_{2}^{(1)} \Phi_{1}^{(1)*}}{\partial \varphi} 
    \left( 2 a^2 \sin^2 \theta + \rho \left[-\Delta' + i a \cos \theta\right] + 2 \Delta \right) \\
 - \frac{2 i a \rho R^2 \sin \theta}{\rho^*} 
    \frac{\partial^2 \Phi_{2}^{(1)} \Phi_{1}^{(1)*}}{\partial t^2} + 2
  \frac{\partial \Phi_{2}^{(1)} \Phi_{1}^{(1)*}}{\partial t} 
    \left( \frac{- \rho R^2 \cot \theta +4 i a^3 \sin \theta }{\rho^*}  + \Delta' \right) + \frac{2 i \Delta \rho \csc \theta}{\rho^*} 
    \frac{\partial^2 \Phi_{2}^{(1)} \Phi_{1}^{(1)*}}{\partial r \partial \varphi}\\ 
    + \frac{2 i a \Delta \rho \sin \theta}{\rho^*} 
    \frac{\partial^2 \Phi_{2}^{(1)} \Phi_{1}^{(1)*}}{\partial t \partial r}+ \frac{\partial \Phi_{2}^{(1)} \Phi_{1}^{(1)*}}{\partial r} 
    \left( \frac{2\Delta ( \rho \cot \theta - 2 i a \sin \theta)}{\rho^*}  \right) - \frac{2 i a \rho \csc \theta}{\rho^*} 
    \frac{\partial^2 \Phi_{2}^{(1)} \Phi_{1}^{(1)*}}{\partial \varphi^2}
 \\
 + \frac{2 a \rho}{\rho^*} 
    \frac{\partial^2 \Phi_{2}^{(1)} \Phi_{1}^{(1)*}}{\partial \theta \partial \varphi} 
     + \frac{2 \rho R^2}{\rho^*} 
    \frac{\partial^2 \Phi_{2}^{(1)} \Phi_{1}^{(1)*}}{\partial t \partial \theta} 
    + \frac{2 (\Delta' \rho - 2 \Delta)}{\rho^*} 
    \frac{\partial \Phi_{2}^{(1)} \Phi_{1}^{(1)*}}{\partial \theta}  - \frac{2 \Delta \rho}{\rho^*} 
    \frac{\partial^2 \Phi_{2}^{(1)} \Phi_{1}^{(1)*}}{\partial r \partial \theta} 
    \\
     + 2 \Phi_{2}^{(1)} \Phi_{1}^{(1)*} 
    \left( \frac{(2 \Delta -  \Delta' \rho) \cot \theta  + 2 i a \Delta' \sin \theta}{\rho^*} 
    - \frac{2 i a \Delta \rho \sin \theta}{\rho^{*3}} \right).
\end{gathered}
\end{equation}

\begin{equation}
\begin{gathered}
\mathcal{B} \, T^{\mathrm{EM}}_{m^*m^*}=  \Delta^2 \left( 
    \frac{5}{\rho^{*2}} 
    - \frac{\rho + 2 r}{\rho^{*3}} 
\right) \Phi_{2}^{(1)} \Phi_0^{(1)*}- \frac{\partial \Phi_{2}^{(1)} \Phi_0^{(1)*}}{\partial t} 
\left( \frac{2 a^2 \Delta r \sin^2 \theta + \Delta \rho R^2}{\rho^{*2}} 
+ \frac{(\Delta' \rho  - 5 \Delta) R^2}{\rho^*} \right) \\
 + \frac{\rho}{\rho^*} \Bigg( 
    - a^2 \frac{\partial^2 \Phi_{2}^{(1)} \Phi_0^{(1)*}}{\partial \varphi^2} 
    - 2 a R^2 \frac{\partial^2 \Phi_{2}^{(1)} \Phi_0^{(1)*}}{\partial t \partial \varphi} 
    + 2 a \Delta \frac{\partial^2 \Phi_{2}^{(1)} \Phi_0^{(1)*}}{\partial r \partial \varphi} 
    - R^4 \frac{\partial^2 \Phi_{2}^{(1)} \Phi_0^{(1)*}}{\partial t^2}  + 2 \Delta R^2 \frac{\partial^2 \Phi_{2}^{(1)} \Phi_0^{(1)*}}{\partial t \partial r} 
    - \Delta^2 \frac{\partial^2 \Phi_{2}^{(1)} \Phi_0^{(1)*}}{\partial r^2} 
\Bigg) \\
 - a \left( \frac{\Delta' \rho - 5 \Delta}{\rho^*} 
    + \frac{\Delta (\rho + 2  r)}{\rho^{*2}} \right) 
    \frac{\partial \Phi_{2}^{(1)} \Phi_0^{(1)*}}{\partial \varphi}  + \Delta^2 \left( \frac{\rho + 2 r}{\rho^{*2}} 
    - \frac{5}{\rho^*} \right) 
    \frac{\partial \Phi_{2}^{(1)} \Phi_0^{(1)*}}{\partial r} \\
\end{gathered}
\end{equation}

\begin{equation}
\begin{gathered}
\mathcal{C} \, T^{\mathrm{EM}}_{nn}=  
+ 2 a^2 \sin^2 \theta \left( \frac{\rho}{\rho^{*3}} + \frac{2}{\rho^{*2}} \right) |\Phi_{2}^{(1)}| 
+ \frac{\rho}{\rho^*} \Bigg( 
    a^2 \sin^2 \theta \frac{\partial^2 |\Phi_{2}^{(1)}|}{\partial t^2} 
    + 2 a \frac{\partial^2 |\Phi_{2}^{(1)}|}{\partial t \partial \varphi} 
   + 2 i a \sin \theta \frac{\partial^2 |\Phi_{2}^{(1)}|}{\partial t \partial \theta} 
\Bigg)  \\
 + \frac{\partial |\Phi_{2}^{(1)}|}{\partial \theta} \left( 
    \frac{\rho \cot \theta - 4 i a \sin \theta}{\rho^*} 
    - \frac{2 i a \rho \sin \theta}{\rho^{*2}} 
\right)  
+ 2\frac{\partial |\Phi_{2}^{(1)}|}{\partial \varphi} \left( 
    -\frac{ a \rho}{\rho^{*2}} 
    - \frac{ 2 a + i \rho \cot \theta \csc \theta}{\rho^*} 
\right)\\ 
+ \frac{\rho}{\rho^*} \Bigg( 
    - \frac{\partial^2 |\Phi_{2}^{(1)}|}{\partial \theta^2} 
    + \csc^2 \theta \frac{\partial^2 |\Phi_{2}^{(1)}|}{\partial \varphi^2}  + 2 i \csc \theta \frac{\partial^2 |\Phi_{2}^{(1)}|}{\partial \varphi \partial \theta} 
\Bigg)-2 a^2 \sin^2 \theta \left( \frac{\rho}{\rho^{*2}} + \frac{2}{\rho^*} \right) 
\frac{\partial |\Phi_{2}^{(1)}|}{\partial t}.
\end{gathered}
\end{equation}

\end{widetext}

As previously mentioned, the source term depends on \(\Phi^{(1)}_{i}\), its first derivative \(\partial_\mu \Phi^{(1)}_{i}\), and various background quantities. Even if the system is not explicitly separable, solving \eqref{phi_0,2 inhomo} remains computationally manageable. However, constructing \(\Phi_1\) can be somewhat more challenging, though it is still far less computationally intensive than reconstructing the metric for the second-order ``pure" gravitational perturbations.

\subsection{Constructing \texorpdfstring{\(\Phi_1\)}{Phi1}}
So far, we have made no specific assumptions about the nature of $\Phi^{(1)}_{i}$ or $\psi^{(1,2)}_{4}$. However, for ringdown analysis or EMRI scenarios-- where the perturber modeled as a point source-- $\Phi^{(1)}_{i}$ can be expressed in a factorized form, which facilitates separating the source. Starting from the homogeneous Maxwell equations, the following ansatz will allow for the separability to the radial and angular Teukolsky equations:
\begin{equation}\label{radial angular Teukolsky eqs}
    \begin{gathered}
        {}_{s} T \, {}_{s} R=0,  \\
        {}_{s} O \, {}_{s} S=0 , \\
   \end{gathered}
\end{equation}
\vspace{1mm}
where ${}_{s} R$ and ${}_{s} S$ are the Teukolsky radial and angular functions and $s=\pm 1$. Subsequently, perturbed Maxwell scalars could be written as 
\begin{widetext}
\begin{equation}
\begin{gathered}
\hat{\Phi}_{2}^{(1)}= {}_{-1} R \, {}_{-1} S \, e^{i \omega t + i m \varphi}\quad\quad\quad
\hat{\Phi}_{0}^{(1)}=  \frac{{}_{+1} R}{\Delta} \, {}_{+1} S \, e^{i \omega t + i m \varphi}\\
\hat{\Phi}_{1}^{(1)}= \frac{1}{4 K Q} \left\{ 
\left[ \left(i \lambda + 2 r \omega - \frac{2 K}{\rho} \right) {}_{-1}R - 
i D \, {}_{+1}R \right] {}_{-1}S+ \left[ i D \, {}_{-1} R - 
\left(i \lambda + 2 r \omega - \frac{2 K}{\rho} \right) {}_{+1}R \right] {}_{+1}S 
\right\}e^{i \omega t + i m \varphi}
\end{gathered}
\end{equation}
\end{widetext}

\begin{figure}
    \centering
    \includegraphics[width=1\columnwidth]{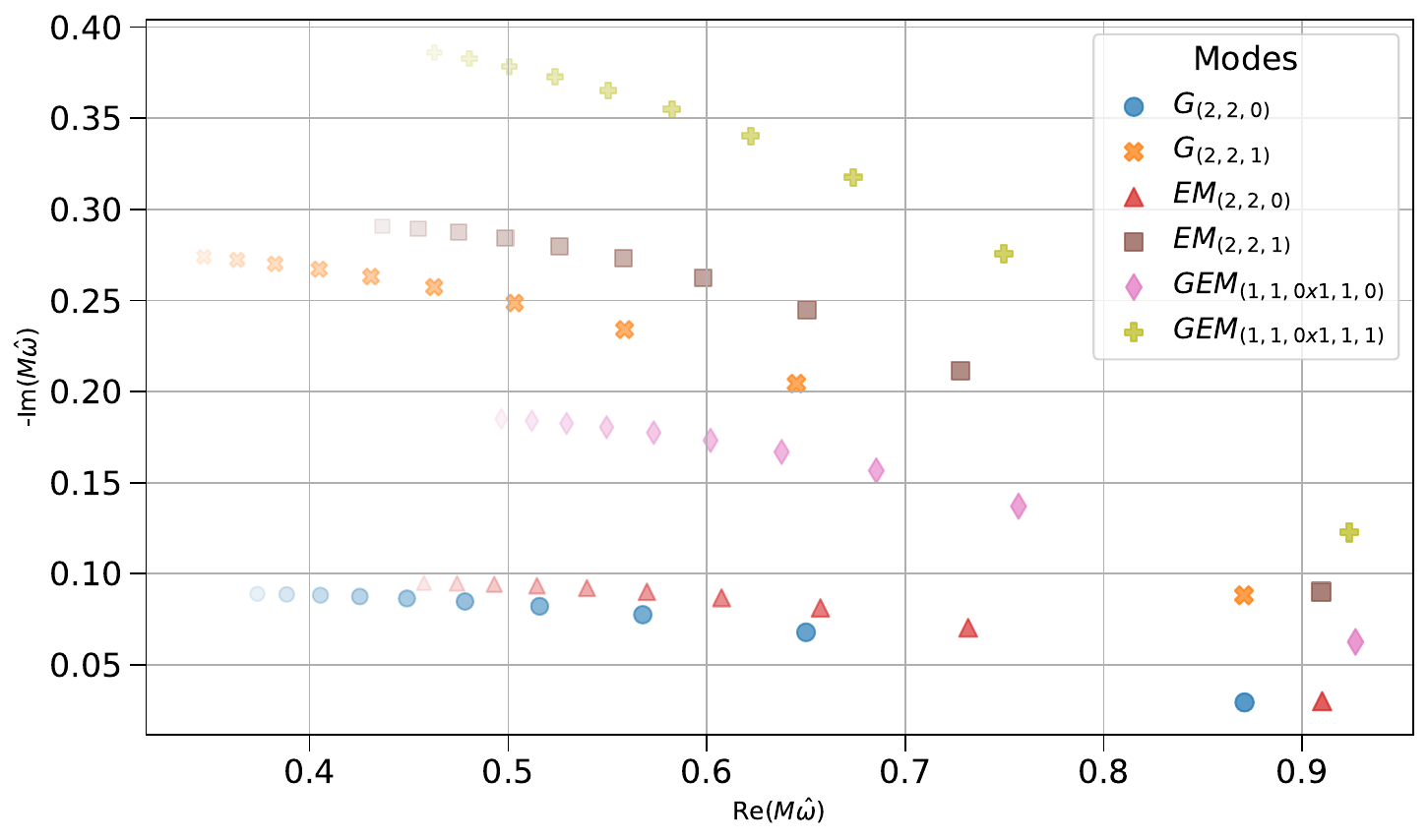}
    \caption{
    Frequencies of linear and quadratic QNMs for various spin values of the final black hole, \( a \), ranging from \( 0 \) to \( 0.99 \) (with 10 evenly spaced values). The modes are categorized into gravitational, GG, and GEM modes: \( G (2,2,0) \), \( G (2,2,1) \), \( GG (1,1,0) \times (1,1,0) \), \( GG (1,1,0) \times (1,1,1) \), \( GEM (1,1,0) \times (1,1,0) \), and \( GEM (1,1,0) \times (1,1,1) \).
    }
    \label{fig:GEM_panel}
\end{figure}

where the hat in \( \hat{\Phi}_{i}^{(1)} \) emphasizes that it represents a solution to the homogeneous system, with the \((l, m, n)\) sub-indexing dropped for convenience.
\begin{equation}
\begin{gathered}
    K = \omega R^2 + a m,\\ 
    Q = a \omega \sin \theta + m \csc \theta,
\end{gathered}
\end{equation}
where \( \lambda \) is the separation constant of the Teukolsky equations \eqref{radial angular Teukolsky eqs}, and \( D \) is the Starobinsky constant \cite{chandrasekhar_1983}.

\subsection{Gravity-Source Separation}
The particular solutions to the cases mentioned earlier are expected to take a separable form. A practical approach involves leveraging the orthonormality of spin-weighted spheroidal harmonics \cite{Pound_2021}.
\begin{equation}
\int{ }_s S_{\ell m}(\theta, \phi ; a \omega)_s \bar{S}_{\ell^{\prime} m^{\prime}}(\theta, \phi ; a \omega) d \Omega=\delta_{\ell \ell^{\prime}} \delta_{m m^{\prime}}
\end{equation}

However, terms involving \(\frac{1}{\rho^{*p}}\) where \(p=1,2,3\), can complicate integration. Under the integration, these terms require a numerical evaluation on a 2D grid of \((r, \theta)\), which might be computationally expensive. Otherwise, the calculation is simpler and can be performed either numerically by integrating only over \(\theta\) or analytically by expanding each spin-weighted spheroidal harmonic $S_{\ell m}$ in terms of spin-weighted spherical harmonics $ Y_{\ell m}$ and use Clebsch-Gordan coefficients (3j-symbols) under nested sums.

To handle the \(\frac{1}{\rho^*}\) dependence systematically, we can also adopt an approach similar to that in \cite{spiers2024analyticallyseparatingsourceteukolsky}. We suggest, however, expanding \(\frac{1}{\rho^*}\) in terms of axisymmetric spherical harmonics:
\begin{equation}\label{rho_spherical}
\frac{1}{\rho^*} = \sum_{l} f_{l}(r) Y_{l0},
\end{equation}
where \(f_{l}(r)\) encodes the radial dependence. This expansion introduces an additional summation over \(l\), but it simplifies the problem by decoupling the angular and radial components. Higher powers of \(\frac{1}{\rho^*}\) can be obtained recursively by taking derivatives of \eqref{rho_spherical}. The resulting series should resemble the known multipole expansions in electrodynamics textbooks.

\section{GEM Coupling}\label{sec:GEM_coupling}

After the coalescence of a BBH, the remnant black hole is believed to relax into a state described by the Kerr solution by emitting GWs characterized by QNMs \cite{ReviewArticle_QNM_Berti_Cardoso,Kokkotas_1999,konoplya2011quasinormal,ferrari2008quasinormal,Pacilio_2018}. Accordingly, at higher orders in perturbation theory, the coupling between gravitational and electromagnetic QNMs is expected to induce a GEM QQNM ``flavor''.\footnote{A more rigorous definition of GEM QQNMs, as poles in frequency space, is provided in complementary work \cite{aly2024nonlinearitieselectromagneticgravitationalmode}.}

Focusing on \(\psi^{\textit{(GEM)}}_4\) and \(\phi^{\textit{(GEM)}}_2\), the governing equations can be written as
\begin{equation}\label{Psi_4_GEM}
    {}_{-2} \mathcal{T} \psi_{4}^{\textit{(GEM)}} = T^{\textit{EM-EM}}\!\left(\phi^{(1)}_{i} \phi^{(1)}_{j}\right),
\end{equation}
where the source term \(T^{\textit{EM-EM}}\) is quadratic in the electromagnetic perturbations and depends on the three first-order Maxwell scalars \(\phi^{(1)}_i\). Similarly,
\begin{equation}\label{phi_2_GEM}
    {}_{-1} \mathcal{T} \phi_{2}^{\textit{(GEM)}} = T^{\textit{G-EM}}\!\left( \phi^{(1)}_{i} \psi^{(1)}_{j} \right),
\end{equation}
where the source term \(T^{\textit{G-EM}}\) includes only the cross-terms of first-order gravitational and electromagnetic perturbations, which requires reconstructing the metric \cite{Ripley_1,Ripley_2,kegeles_cohen_1979,chrzanowski_1975} along with the Maxwell tensor. In \eqref{Psi_4_GEM} and \eqref{phi_2_GEM}, the operator \({}_{s} \mathcal{T}\) denotes the Teukolsky operator for spin weight \(s\).

In \cite{Fawzi_EM_idealdipole,aly2024nonlinearitieselectromagneticgravitationalmode}, the GEM QQNM frequencies for the gravitational and electromagnetic sectors were found to be
\begin{equation}\label{Psi_4_GEM_QNM}
\begin{aligned}
   \omega^{\textit{(GEM)}}_{(l_{1},m_{1},n_{1}) \times (l_{2},m_{2},n_{2})}
   &= \Omega_{(l_{1},m_{1},n_{1})} \pm \Omega_{(l_{2},m_{2},n_{2})},
\end{aligned}
\end{equation}
\begin{equation}\label{phi_2_GEM_QNM}
\begin{aligned}
    \Omega^{\textit{(GEM)}}_{(l_{1},m_{1},n_{1}) \times (l_{2},m_{2},n_{2})}
    &= \Omega_{(l_{1},m_{1},n_{1})} \pm \omega_{(l_{2},m_{2},n_{2})},
\end{aligned}
\end{equation}
where \(\omega_{(l,m,n)}\) and \(\Omega_{(l,m,n)}\) represent the gravitational and electromagnetic QNM frequencies, respectively.

GEM QQNMs are expected to have longer lifetimes compared to the first overtone in each sector, as exemplified for the \((l=4,m=0)\) and \((l=2,m=0)\) harmonics in Fig.~\ref{fig:GEM_panel}. This behavior is qualitatively similar to that of GG QQNMs, reflecting the similarity of electromagnetic and gravitational QNMs.

Moreover, GEM QQNMs are not solely dictated by their parent modes, but also by the underlying coupling between the electromagnetic and gravitational sectors. In our minimally coupled Einstein--Maxwell setup this coupling is fixed by the standard stress--energy source terms entering Eqs.~\eqref{Psi_4_GEM_QNM} and \eqref{phi_2_GEM_QNM}. In this sense, the GEM amplitudes in this framework probe essentially the same interaction structure that sources the linear electromagnetic and gravitational modes, so any departure from minimal coupling would manifest itself as a deformation of the associated spectra.

A useful example of this sensitivity is provided by Einstein--Maxwell--dilaton (EMD) theory \cite{Pacilio_2018}. Pacilio and Brito computed the gravitational and electromagnetic QNM spectra of weakly charged, static and slowly rotating EMD black holes as functions of the dilaton coupling \(\eta\) and charge-to-mass ratio \(v = Q/M\). They found that, in the static case, the fundamental gravitational QNMs depend only weakly on \(\eta\) (with relative shifts at the \(\lesssim \mathcal{O}(10^{-2})\) level across \(0 \lesssim \eta \lesssim 3\)), whereas the electromagnetic spectrum exhibits a much stronger dilaton dependence and a pronounced breaking of axial/polar isospectrality, reaching \(\sim 10\%\) differences in the polar sector for \(v \sim 0.6\) \cite{Pacilio_2018}.

\begin{figure}
    \centering
    \includegraphics[width=1\columnwidth]{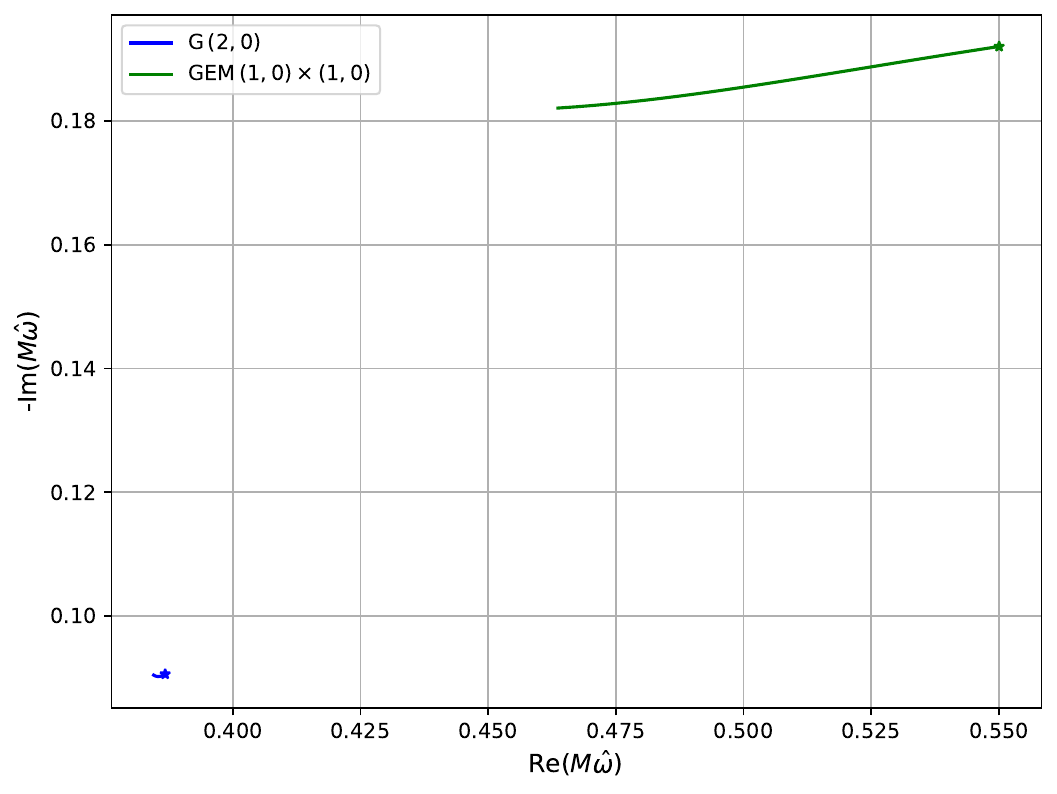}
    \caption{
    Frequencies of the polar gravitational QNM \((2,0)\) and the polar GEM QQNM \((1,0)\times(1,0)\) on a static charged background with charge-to-mass ratio \(v = Q/M = 0.6\) in Einstein--Maxwell--dilaton theory, shown as functions of the dilaton coupling \(\eta \in [0,1.52]\). The stars denote the corresponding QNMs predicted by minimally coupled Einstein--Maxwell theory at \(\eta = 0\) for both modes. The electromagnetic sector exhibits a much stronger \(\eta\)-dependence than the gravitational one, so that the GEM quadratic mode inherits an enhanced sensitivity to the dilaton coupling at the level of the spectrum. This figure is intended as a proof of concept for the coupling dependence of GEM QQNMs; whether such differences are observationally accessible depends on the achievable signal-to-noise ratio, detector sensitivity, and systematic uncertainties, and we do not attempt a detailed forecast here. The data used in this figure are taken from \cite{Pacilio_2018,QNM_EMD_notebook}.
    }
    \label{minimal coupling}
\end{figure}

In Fig.~\ref{minimal coupling} we exploit their tabulated static QNMs for \(v = Q/M = 0.6\) to construct the corresponding polar GEM QQNM \((1,0)\times(1,0)\) from the quadratic combination of the parent polar electromagnetic mode \((1,0)\), and to compare it with the ordinary polar gravitational mode \((2,0)\), both shown as functions of the dilaton coupling \(\eta\). While the gravitational mode frequency varies only mildly with \(\eta\), the electromagnetic parent mode is significantly more \(\eta\)-sensitive, so that its quadratic GEM descendant inherits an even larger fractional variation than the linear gravitational spectrum. This proof-of-concept example suggests that, in EMD-like theories and, more generally, in scalar–tensor or tensor–vector–scalar extensions where the electromagnetic sector is coupled to additional fields, GEM QQNMs can be parametrically more sensitive to the coupling than both linear gravitational QNMs and GG quadratic modes at the level of the underlying spectrum.

This pattern is broadly consistent with other families of gravity–electromagnetism couplings studied in the literature. Nonminimal Einstein--Maxwell models with curvature couplings of the schematic form \(R F^{2}\) or \(R_{\mu\nu}F^{\mu\alpha}F^{\nu}{}_{\alpha}\) modify the propagation of electromagnetic fields on curved backgrounds and are expected to deform the electromagnetic-led QNM spectra more strongly than the tensor sector \cite{Balakin_Lemos_2005,Dereli_Sert_2011}. Similarly, in Einstein--Maxwell–scalar theories with couplings \(f(\phi)F^{2}\), scalarized charged black holes develop mixed scalar/electromagnetic/gravitational modes whose frequencies and damping times can be highly sensitive to the scalar–Maxwell coupling function \cite{Myung_Zou_2019,BlazquezSalcedo_2021,Guo_2023}, and in which axial–polar isospectrality is generically broken in the electromagnetic and gravitational sectors \cite{BlazquezSalcedo_2021}. In all these cases the electromagnetic (or scalar–electromagnetic) channels typically respond more strongly to changes in the coupling than the purely tensorial sector.

\section{Discussion}\label{Discussion}
 In the context of the ringdown analysis, the situation differs slightly from EMRIs, as defining a perturbation parameter is not straightforward in general \cite{Gleiser_Second_Schwarzschild1996_CloseApprox}. Yet, numerical simulations of head-on or quasi-circular inspirals of charged BBHs \cite{Zilhao_equal_Charge,Zilhao_equal_opposite_Charge,Cardoso_minichargedDM,Liebling_2016,Bozzola_2021,Bozzola_2021_partII,Luna_2023} highlight that both gravitational and electromagnetic radiation are emitted during all phases of coalescence. The ringdown phase, as expected, is characterized by QNM frequencies from both types of radiation. Depending on the binary's parameters, electromagnetic contributions can represent a significant fraction of the total emission, underscoring their importance in black hole spectroscopy for this category of mergers\cite{ReviewArticle_QNM_Berti_Cardoso,2Modes_As_test_Per_theory_1,2Modes_As_test_Per_theory_2_Agnosticspectroscopy}. 
\subsection{Charged black hole mergers}
Charged black hole mergers can involve two like-charged holes, one charged and one neutral, or oppositely charged components. The charge configuration critically shapes the dynamics and radiation of the remnant. In the oppositely charged case, conservation laws imply a net charge \( Q_{\text{tot}} = ||Q_1| - |Q_2|| \), and energy is radiated primarily through low multipole modes, notably the dipole \((\ell=1, m=1)\). In such scenarios, electromagnetic radiation can rival or exceed gravitational radiation, making nonlinear GEM interactions as relevant as GG. Indeed, for head-on collisions, it was shown in \cite{Zilhao_equal_opposite_Charge} that EM dipole emission becomes comparable to gravitational waves for \( |Q|/M \gtrsim 0.2 \), and even dominates for \( |Q|/M \gtrsim 0.37 \). Similar behavior is expected in more general merger configurations.

To investigate nonlinear effects in numerical simulations, one typically uses envelope-based estimates to assess the plausibility of identifying quadratic signatures. For example, one can compare the noise level in waveforms—extracted at null or spatial infinity—to the square of the parent mode waveform, in order to judge whether QQNMs could manifest. This approach applies to both GG and GEM interactions. Moreover, if gravitational radiation is expected to dominate certain harmonics, we can directly assess whether GEM contributions are comparable to those of GG interaction by comparing the energy radiated in both electromagnetic and gravitational channels. This offers a first-step, practical criterion for determining whether GEM can be meaningfully extracted against the noise estimated in the child gravity mode. Numerous studies \cite{Redondo_Yuste_2024,spin_denpdence_Hengrui_Zhu,Ma_2024,Bucciotti_2024,Bucciotti_2023,bourg2024quadraticquasinormalmodedependence,Nakano_2007} have explored the correlation between the amplitudes of GG QQNMs, \(\psi_{4}^{(2)}\), and their parent mode amplitudes, \(\psi_{4}^{(1)}\), as well as how this relationship can depend on spin, initial data, and mode parity. This correlation is typically expressed through the ratio:

\begin{equation}
    R^{GG}_{(l_{1}, m_{1}, n_{1}) \times (l_{2}, m_{2}, n_{2})} = 
    \frac{A^{(2)}_{(l_{1}, m_{1}, n_{1}) \times (l_{2}, m_{2}, n_{2})}}{A^{(1)}_{(l_{1}, m_{1}, n_{1})} A^{(1)}_{(l_{2}, m_{2}, n_{2})}},
\end{equation}
where \( A^{(1)}_{(n,l,m)} \) and \( A^{(2)}_{(l_{1},m_{1},n_{1}) \times (l_{2},m_{2},n_{2})} \) represent the amplitudes of the parent and child modes, respectively, due to GG interactions. Accordingly, for GEM modes, we define the corresponding ratio
\begin{equation}
    R^{GEM}_{(l_{1}, m_{1}, n_{1}) \times (l_{2}, m_{2}, n_{2})} = 
    \frac{A^{(GEM)}_{(l_{1}, m_{1}, n_{1}) \times (l_{2}, m_{2}, n_{2})}}{B^{(1)}_{(l_{1}, m_{1}, n_{1})} B^{(1)}_{(l_{2}, m_{2}, n_{2})}},
\end{equation}
where \( B^{(1)}_{(n,l,m)} \) denotes the electromagnetic mode amplitude, and \( A^{(GEM)}_{(l_{1},m_{1},n_{1}) \times (l_{2},m_{2},n_{2})} \) represents the amplitude of the child mode arising from GEM interactions.

Since both \( R^{GEM}\) and \( R^{GG}\) are expected to depend on the background details, and maybe also on other factors, we defer quantifying the ratio between the background factors to future work and instead we focus here on the ratio between parent modes
\begin{equation}\label{gamma_factor}
\begin{aligned}
    \Gamma^{(l'_{1}, m'_{1}, n'_{1}) \times (l'_{2}, m'_{2}, n'_{2})}_{(l_{1}, m_{1}, n_{1}) \times (l_{2}, m_{2}, n_{2})} &\equiv 
    \frac{A^{(GEM)}_{(l'_{1}, m'_{1}, n'_{1}) \times (l'_{2}, m'_{2}, n'_{2})}}{A^{(2)}_{(l_{1}, m_{1}, n_{1}) \times (l_{2}, m_{2}, n_{2})}} \\ 
    &\sim 
    \frac{B^{(1)}_{(l'_{1}, m'_{1}, n'_{1})} B^{(1)}_{(l'_{2}, m'_{2}, n'_{2})}}{A^{(1)}_{(l_{1}, m_{1}, n_{1})} A^{(1)}_{(l_{2}, m_{2}, n_{2})}}.
\end{aligned}
\end{equation}

To leading order at infinity, 
\begin{equation}
\begin{gathered}
    \psi_{lmn\,4}^{(1)} \simeq A_{lmn}^{(1)} {}_{-2}S_{l m}\,
 e^{i\omega (t-r_*) + i m\varphi},\\
 \phi_{lmn\,2}^{(1)} \simeq B_{lmn}^{(1)} {}_{-1}S_{l m}\,
 e^{i\omega (t-r_*) + i m\varphi},
 \end{gathered}
\end{equation}   
with the spin–weighted spheroidal harmonics normalized. The gravitational and electromagnetic powers carried by this single mode are
\begin{equation}
P_{\rm GW}^{}
 \;=\;
 \frac{|B^{(1)}|^2}{4\pi\,|\omega|^2},
\qquad P_{ EM}^{}
 \;=\;
 \frac{|A^{(1)}|^2}{4\pi}.
\end{equation}

as we focusing on dominate mode of each chanel, the by suppressing the \( lmn \) indices, 
\begin{equation}\label{gamma_simplified}
    \Gamma \sim \frac{B^{(1)2}}{A^{(1)2}} = |\omega|^2 \frac{\dot{E}^{RD}_{EM}}{\dot{E}^{RD}_{GW}},
\end{equation}
 where \( \dot{E}^{RD} \) is the rate of energy emission during the ringdown phase. For instance, in the opposite charge binaries, $\Gamma$ magnitude should be greater than in same charge systems.    
\vspace{1mm}

Moreover, for a head-on collision between equal-mass, equal-charge-ratio charged BBHs in \cite{Zilhao_equal_Charge}, the ratio between the dominant modes from each sector was reported to be approximately \(\Gamma \sim 0.1 \), and observed around the frequencies of the corresponding \((2, 0, 0)\)-gravity QNMs modes. 

Also in \cite{Zilhao_equal_Charge}, the electromagnetic and gravitational fluxes share similar time profiles and comparable magnitudes across various charge values \( Q \), with their ratio approximately following \( E_{EM}/E_{GW} \sim Q^2 / 4M^2 \). This suggests that, when the signal-to-noise ratio is sufficiently high, GEM signatures could be detectable under the same conditions where GG signals are already identifiable.

Furthermore, for the case of semi-circular inspirals \cite{Bozzola_2021,Bozzola_2021_partII}, the ratio \( E_{EM}/E_{GW} \) during the inspiral phase up to the merger point was estimated to be of the order of \( 10^{-1} \) for \( |Q|/M \gtrsim 0.2 \), with the values increasing for higher charge-to-mass ratios. The same holds for \( J_{EM}/J_{GW} \), which is quantitatively related to the ratio \( \Gamma \) in \eqref{gamma_factor}.

Finally, we emphasize that in the case of charged BBH mergers, the remnant black hole is typically born charged. As a result, the problem must be reformulated on a charged background, such as Kerr–Newman \cite{Mark_2015,Berti_2005,Pani__QNM_Slow_KN} or Reissner–Nordström \cite{chandrasekhar_1983,griffiths_podolský_2012,Aly_2024}. Nevertheless, the analysis is expected to remain similar to that in the Kerr case.
\subsection{magnetized systems}
Scenarios involving BNS mergers, magnetar collapse, and NSBH mergers without significant disruption are of greater astrophysical relevance and serve as promising laboratories to search for GEM QQNMs in the ringdown phase of their remnant black holes. In certain BNS scenarios, a hypermassive neutron star forms temporarily before collapsing into a final black hole. This transient NS typically acquires extremely high angular momentum, generating magnetic fields that may reach up to \( B \sim 10^{12} \, \text{T} \), as is believed to have occurred in GW170817 \cite{abbott2017gw170817}.

This raises intriguing questions about the role of electromagnetic fields in such extreme conditions and whether they can leave detectable imprints on gravitational signals. Current and future simulations of magnetized systems could provide valuable insights into the remnant’s ringdown for these scenarios, particularly in light of recent efforts to underscore electromagnetic QNMs, as discussed in \cite{Most_2018,Most_2024}. However, the presence of a non-vacuum background complicates the application of state-of-the-art methodologies commonly used in ringdown analysis. Nevertheless, with extremely strong magnetic fields, it is worth exploring whether nonlinearities could be identified in such simulations, particularly near the horizon, similar to the shear mode analysis in \cite{Mourier_2021,NearHorizonNonlinearities}.

 Qualitatively, GEM effects can emerge during the inspiral phase of an EMRI due to the energy content of the electromagnetic field
. Sophisticated calculations within the framework of post-Newtonian formalism or black hole perturbation theory could provide hints whether GEM effects could be detected in signals relevant to the upcoming LISA mission. We aim to address such calculations in future work.

\section{Conclusion}\label{Conclusion}

In this work we have extended previous studies of GEM coupling on Schwarzschild backgrounds to the more astrophysically relevant case of Kerr spacetime. Within Teukolsky formalisms, we derived explicit quadratic electromagnetic sources for the gravitational Teukolsky equation and re-visited how GEM interactions generate a new family of GEM QQNMs in both the gravitational and electromagnetic sectors. These modes enrich the spectrum beyond purely gravitational quadratic QNMs and provide a concrete framework in which to organize nonlinear GEM effects in rotating black hole spacetimes. For the charged BBH mergers, we expect the remannt to be charged hence, next stop in the analytical research should be the Kerr-Newman spacetime.

We then examined several astrophysical situations in which GEM signatures may become comparable to, or even rival, purely gravitational nonlinearities. For charged BBH mergers, existing simulations indicate that electromagnetic radiation can contribute a non-negligible fraction of the radiated energy, especially in oppositely charged configurations, suggesting that GEM QQNMs could be detectable under the same conditions where quadratic gravitational modes are already accessible. In magnetized mergers such as BNS or NSBH systems, the presence of ultra-strong magnetic fields, potentially as large as \( B \sim 10^{12} \,\mathrm{T} \), raises the prospect that GEM effects may leave measurable imprints in the ringdown of the remnant, although the non-vacuum nature of these environments complicates the analysis.

Beyond waveform systematics, GEM modes open a number of conceptual and phenomenological opportunities. Because any form of energy gravitates, GEM signatures provide a natural handle on hidden sectors, including black holes charged under a dark \( U(1) \) gauge symmetry that is otherwise decoupled from the Standard Model. Moreover, modifying the coupling between gravity and electromagnetism---for instance in Maxwell–dilaton–Einstein or scalar–vector–tensor theories---alters the GEM QQNM spectrum and offers a concrete way to test the minimal coupling principle. While magnetized mergers are more common in nature, charged BBH mergers constitute a particularly clean theoretical laboratory in which to benchmark GEM effects. 

Finally, although we have focused on Kerr backgrounds as a clean starting point, charged mergers naturally point toward Kerr--Newman geometry \cite{Mark_2015,Berti_2005,Pani__QNM_Slow_KN,chandrasekhar_1983,griffiths_podolský_2012,Aly_2024}. A natural next step is therefore to generalize the GEM analysis to the later one, implement it in perturbative solvers and numerical-relativity frameworks, and apply it to existing and future simulations of charged and magnetized systems. Mapping out the regions of parameter space where GEM signatures become comparable to GG nonlinearities, and assessing under what conditions they might be constrained with ground-based detectors and LISA, are key goals for future work.

\section{Acknowledgments.}
We extend our deepest gratitude to Mahmoud A. Mansour for his invaluable time, support, and assistance throughout this project. We also sincerely appreciate the insightful conversations with Jacob Fields and Jaime Redondo Yuste. Additionally, we are grateful to Vasileios Paschalidis, David Radice, and Elias Most for their valuable discussions. Finally, we acknowledge the partial support of D.S. through the U.S. National Science Foundation under Grant No. PHY-2310363.

\bibliography{main}
\end{document}